\documentclass[10pt]{article}
\pdfoutput=1
\usepackage[utf8]{inputenc}
\usepackage{graphicx}
\usepackage{courier}
\usepackage{longtable}
\usepackage{tabulary}
\usepackage{booktabs,array,multirow}
\usepackage[greek,english]{babel}
\usepackage[labelfont=bf]{caption}
\usepackage{authblk}
\usepackage{dblfloatfix}
\usepackage{fancyhdr}
\usepackage{geometry}

\title{LiFi: Enlightening Communications}
\author{Christophe Jurczak}
\affil{\small\texttt{christophe.jurczak@lucibel.com}}

\date{\today}

\begin{document}

\captionsetup[figure]{labelfont={bf},name={Figure}}
\captionsetup[longtable]{labelfont={bf},name={Table}}

\maketitle

\begin{abstract}
LiFi is a networked wireless communication technology transforming solid-state indoor lighting into a backbone for information. The technology has reached maturity, with the first LiFi LED luminaire commercialized in 2016 by Lucibel and PureLiFi. More than 50 clients of Lucibel have built projects with a large variety of use cases. A very high connection density as well as staggering bandwidth improvements in the lab in excess of 10 Gbps, hint to a luminous future for LiFi as a powerful complement or in certains situations an alternative to WiFi and 4G/5G.
\end{abstract}

\section{Lighting and LiFi}
Among the major innovations in the energy sector, the development of LED lighting has been nothing less than earth shattering over the last 10 years, largely thanks to the virtuous conjunction of technology development and policies and measures in support of energy efficiency and renewable energy. Common LED bulbs today consume 85\% less energy than their incandescent counterparts and their deployment is poised to have a massive impact on the energy mix, with lighting accounting currently for as much as 15\% of global electricity consumption and 5\% of worldwide greenhouse gas emissions. 

Costs have been slashed down through improvements in manufacturing and higher wall-plug efficiency for a given optic illumination. Costs per LED bulb remain higher than for incandescent and fluorescent technologies but a much longer lifespan up to 25,000 hours and energy savings make the cost per lumen – a measure of the total quantity of visible light emitted by a source – much more competitive. 
As a consequence, the LED lighting market share is already 40-50\% (depending on the geography) and 70\% of a global \$100bn general lighting market will stem from LED shipments (lamps and luminaires) as early as 2020. In the US alone, LED installed stock is expected to grow from 6\% in 2016 to close to 60\% in 2025 and 90\% in 2035 \cite{NAS2}. 

While the lighting industry has been traditionally slow to adapt and move, with product cycles longer than 10 years, LED lighting is providing the opportunity, at an accelerated pace, to improve existing business operations on the one hand, and create new business opportunities on the other hand. With the Solid-State Lighting (SSL) technologies - including LED, Organic LED (OLED) and Laser Diode technologies - come lighting control and communication systems that bring intelligence at the level of the AC-to-DC driver. These are greatly expanding the functions that can be performed by a luminaire, making it a central part of the Smart Building and Smart City concepts. 

\begin{figure}
  \includegraphics[width=\linewidth]{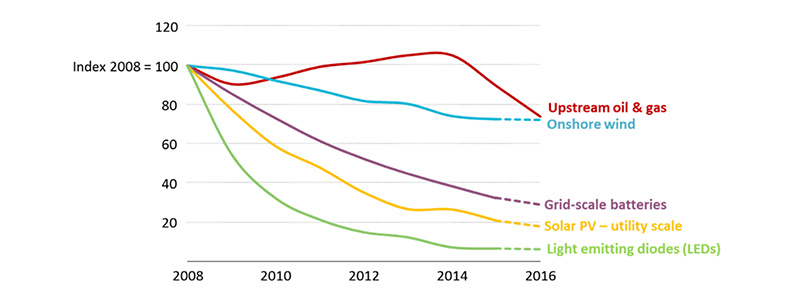}
  \caption{Recent cost trends for selected energy technologies. The LED sector has outperformed over the last 10 years all energy technologies, including solar
photovoltaic. Source: World Energy Outlook 2016, IEA}
  \label{fig:weo}
\end{figure}

Lucibel has been since 2008 at the forefront of the LED revolution, manufacturing and distributing luminaires and offering innovative indoor lighting solutions for its Commercial and Industrial customers. While downward trending prices and commoditization are happening and re-shaping the industry, Lucibel is always looking for ways to provide groundbreaking value-adding features to its customers. Control systems, connected or not, act only on one of the two components of light, its intensity, in a conventional, quasi-static way. Lucibel builds on these systems and adds new functionalities to address two pressing needs of our society: well-being and high-speed communications.

The first development is to program or adjust on demand the spectral content of a LED luminaire ({\it i.e.} its color) to follow circadian rhythms in the course of the day in a biologically and emotionally effective way. The second concept is to modulate the intensity at a flicker-free high frequency to code information into light and create a very high-speed wireless data exchange channel between a luminaire and a receiver.  This technology is termed as Visible Light Communication (VLC) and complete wireless networking using VLC is called LiFi, a term coined by Professor Haas – also co-founder of Lucibel’s technology partner PureLiFi - at his 2011 TED Global Talk \cite{haasted2011} where he introduced the idea of “wireless data from every light”. 

Optical communications are nothing new. First developed in the 1970s, fiber-optics have revolutionized the telecommunications industry and have been a major enabler of the current Information Age. The internet runs mostly on optical fibers and last-mile fibers are more and more common. LiFi extends the optical communication revolution closer to the final customer, to the last meter, transforming indoor lighting into a backbone for information. 

\section{Light Communications}

There are multiple ways to encode information into light, and they all involve some combination of amplitude, frequency and phase modulation (or keying in the language of the digital signals community). The ultimate goal of the signal processing is to achieve reliable wireless communications with a minimal bit-error ratio between any two nodes in the network, {\it i.e.} a given ratio of bits in error relative to the total number of transmitted bits, as a function of the signal to noise ratio of the modulated signal. Bit errors occur over a communication channel because of noise, interference, distortion or synchronization issues.

\begin{figure}[h]
  \includegraphics[width=\linewidth]{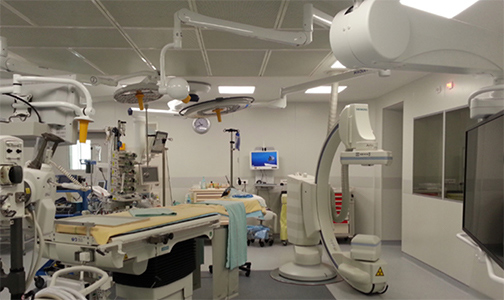}
  \caption{Short term applications for LiFi are in places where high-speed wireless connectivity has to be guaranteed and secured or where RF communications are hampered or prohibited, for example in hospitals. Source: APHP}
  \label{fig:hopital}
\end{figure}

In the context of VLC, light is emitted by a LED and detected either by a single photodetector (a photodiode or an avalanche photodiode in case of low irradiation) or an imaging sensor, for example the camera of a consumer electronic device. In this last case, one talks about Optical Camera Communication (OCC). With such a pair of front end components, VLC poses a tricky challenge: contrary to what’s happening in the Radio Frequency (RF) domain, the amplitude and phase of the electromagnetic field emitted by LEDs can’t be separately modulated and detected because of the absence of a reference local oscillator at the detection point, and data transmission is only doable as an intensity modulation and direct detection scheme. This imposes a constraint on the signal that can be used to modulate the LED through the current driver: it has to be real valued and strictly positive to be successfully mapped into the light intensity and that limits the typology of eligible modulation schemes with respect to the RF domain.

\begin{figure}[t!]
  \includegraphics[width=\linewidth]{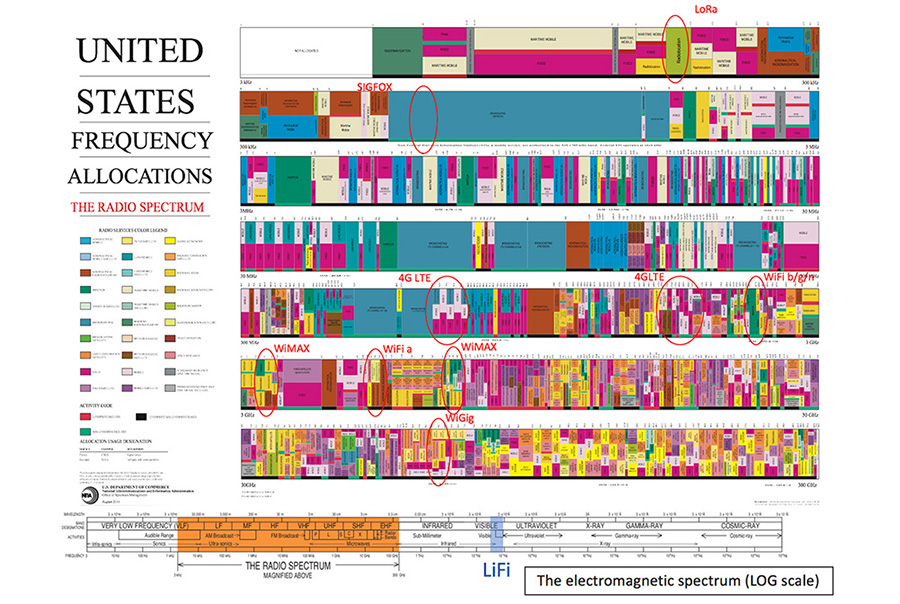}
  \caption{US frequency allocation chart (logarithmic scale). The RF spectrum spans the 3kHz-300GHz range of the electromagnetic spectrum and is heavily regulated and used. The most common bands for consumer electronics and IoT are shown on the graph. By contrast the visible light and near infrared spectrum from 250 THz to 800 THz, more usually expressed in wavelength from 400 nm to 1.2 \selectlanguage{greek}u\selectlanguage{english}m, is not regulated and can be used for light communications over a much larger range of frequencies. Sources: US Department of Commerce \cite{freq}, Lucibel}
  \label{fig:frequencychart}
\end{figure}

On the flipside, the LiFi domain, which is the near IR and visible light frequency range, is not regulated and is orders a magnitude larger than the RF range, suffering from the “spectrum crunch” due to the exponentially increasing need for bandwidth.

Finally, the modulation scheme must be designed with the characteristics of the luminaire in mind: dimming must be possible, no flickering should be perceived (achieved with a switching rate faster than 2kHz), the LED visual appearance characterized by the Color Rendering Index (CRI) and Correlated Color Temperature (CCT) shouldn’t be impacted, power loss and heat should be minimized. 

Indeed, LiFi systems must be designed as illumination systems with communications capabilities, not the reverse.

\section{LED Modulation Techniques}

Single Carrier Modulation (SCM) techniques are relatively straightforward to implement in Light Communications. Illumination control can be supported by adjusting the light intensities of the “on” and “off” states at typically +/- 10\% of dimming level, without affecting the system performance. Different kinds of pulses (symbols) are used, such as Non-Return-to-Zero or Manchester codes.

On-Off Keying (OOK) is the technology of choice for OCC and applications such as location based marketing where a limited amount of contextual information is pushed to a mobile device. But, for high speed communiactions, performances deteriorate as the bit rates increase. Sophisticated equalization techniques with different degrees of performance and computational complexity, as well as significant power consumption, are required for SCM techniques to be effective at high data rates which makes them unattractive. The IEEE Standards Association has developed the 802.15.7 standard \cite{ieee15} for this kind of short-range and low data rate communication using visible light.

\begin{figure}[h]
  \includegraphics[width=\linewidth]{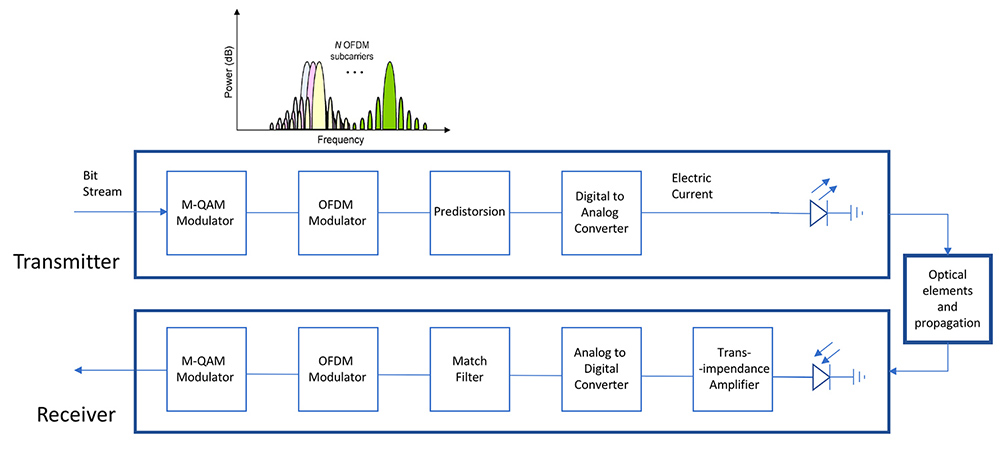}
  \caption{Block Diagram of a high speed LiFi device. Through the allocation of signal bits among sub-carriers, OFDM is a technique that allows to pack much more signal in a given frequency carrier than Single Carrier Modulation techniques. The subcarrier frequencies are chosen so that the signals are mathematically orthogonal over one symbol period. Pre-distortion is used to linearize the dynamic range of the LED. Adapted from \cite{hass2012li}}
  \label{fig:ofdm}
\end{figure}

For high-speed optical wireless communication, Multi-Carrier Modulation (MCM) is preferred. For example, PureLiFi’s technology incorporated into Lucibel's devices relies on Orthogonal Frequency Division Multiplexing (OFDM) \cite{armstrong2009ofdm}\cite{afgani2006visible}, where parallel data streams are transmitted simultaneously through a collection of orthogonal subcarriers. The spectra of individual subcarriers overlap, but because of the orthogonality property the subcarriers can be demodulated without interference. The subcarrier spacing is the inverse of the symbol duration and the subcarrier bandwidth is designed so that it is smaller than the channel coherence bandwidth in order to omit complex equalizer circuitry.

At the level of the LED, Quadrature Amplitude Modulation (QAM) is used to encode data into symbols loaded afterwards into the subcarriers. QAM introduces complexity but is more robust to noise than alternative schemes at high speeds. The parallel symbols can then be multiplexed into a serial time domain output using Inverse Fast Fourier Transformation (IFFT) and then de-multiplexed after using FFT at the receiver. A so called Cyclic Prefix is added to the start of each time domain OFDM symbol before transmission, this way eliminating both inter-symbol and inter-channel interference from the received signal. 

OFDM is a well-known technique used in RF communication protocols for WiFi, powerline and 4G communications and its implementation for light communication is almost equivalent but for one important characteristic: RF OFDM generates complex-valued negative and positive (bipolar) signals, incompatible with LED intensity modulation. With a proper symmetry operation, the time domain signal is forced into the real domain but at the detriment of a reduction of the system bandwidth by a half. A simple way to make a bipolar signal strictly positive is to introduce a direct current (DC) bias around which the original bipolar signal can vary. This scheme is known as DC-biased Optical OFDM (DCO-OFDM). Numerous variant OFDM techniques have been developed over the last 10 years to provide energy efficient alternatives without the sacrifice of spectral efficiency. This is still a subject of intense research \cite{islim2016modulation}\cite{noshad2016hadamard}. 

As well as its many advantages, OFDM has a number of disadvantages, such as the high peak-to-average power ratio which imposes a wide dynamic range in many of the components of the transmitter and the receiver. Very fast Digital Signal Processing implementation is required to perform Fourier Transforms and the design of the Digital-Analog converters is critical because of the complexity of the signal and the required accuracy of the conversion. All these requirements entail a cost but this technology scales up quickly, in a similar way to WiFi components.

\section{LiFi as a Communication Solution }

LiFi is not only a photonic virtual cord, it is a complete wireless networking system, offering bi-directional multi-user communication, within a wireless network of very small optical cells,  therefore a very high spatial connection density, and with seamless handover. Each LiFi luminaire is an Access Point (AP) \cite{dimitrov2015principles}.

Optical OFDM provides natively a multiple access technique called OFDMA, also the method of access for the new WiFiax standard, where users of data broadcasted by a given luminaire are separated by a number of orthogonal subcarriers. Other multi-user access technologies can also be used such as Time Division Multiple Access (TDMA).  

For a complete LiFi communication system, duplex communication is required, {\it i.e.} an uplink connection from the mobile terminals to the optical AP should be provided. RF duplex techniques where the downlink and the uplink are separated by different time slots, or different frequency bands, could be used. However, emitting intense white light by the receiver terminal is not acceptable in practice. A solution, implemented by Lucibel and PureLiFi, is to use Wavelength Division Duplexing (WDD), using visible light modulation for the downlink and the modulation of an IR LED for the uplink communication channel. Using RF communication for the uplink is also an option in certain configurations since there is often a traffic imbalance in current wireless communication systems that makes the uplink channel considerably less congested.

In RF wireless communications, the network is distributed over areas called cells, each served by at least one fixed-location base station. In 4G LTE, in order to improve user access, the network is densified by the addition of cells of different sizes referred to as macro-, micro-, pico- and femto-cells in order of decreasing base station power. Inter- and intra-cell interference avoidance is one of the most critical challenges for the concurrent operation of these cells. 5G wireless network should see the incorporation of unlicensed networks such as WiFi into so called heterogeneous networks. 

The concept of cell is easily transposed to LiFi and the optical AP associated to a LiFi luminaire is frequently called an “attocell” because of its small size. Because of the density of luminaires and the nearly uniform illuminance, optical attocells can drastically improve coverage and data density. In an hybrid LiFi/WiFi deployment \cite{ayyash2016coexistence}, LiFi offloads some of the traffic from WiFi to maintain the broadband user experience. 

With LiFi, each user benefits from the bandwidth available under each luminaire, without sharing it with users under other luminaires. This idea of "connection densification" is a key point in favor of LiFi with respect to WiFi, as much as the high data rate.

Moreover, whereas complex beamforming techniques are developed for next generation RF wireless systems to increase capacity, beamforming is native for light based communications. It is worth noting that, similar to conventional RF based communication systems, Multiple Input-Multiple Output (MIMO) schemes in LiFi and more generally VLC are capable of bringing data transmission speed enhancements \cite{zhang2017localisation}. 

All these characteristics make LiFi a crucial enabler of mobility indoor with as yet unmatched quality of service, improving substantially the user experience.

A Study Group of the IEEE Standards Association is exploring adding light communication protocols to the 802.11 WiFi standards for communications \cite{ieeelcsg}\cite{burchardt2014vlc}. This standardisation effort is an important factor towards large scale deployments of LiFi and Lucibel is a contributor to the work being performed.

\begin{figure}[b!]
  \includegraphics[width=\linewidth]{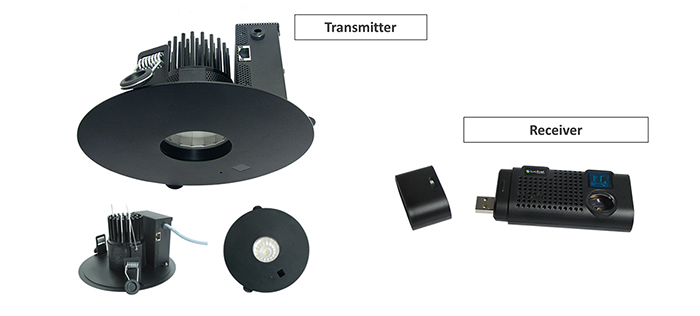}
  \caption{The first commercial LiFi downlight, manufactured by Lucibel and PureLiFi and brought to the market in September 2016. Rated performance is 42 Mbps downlink and uplink in a standard lighting geometry at a height of 2.5 m. 8 users can be served data simultaneously by a luminaire. The uplink channel is in the near infrared band to avoid glare from the USB receiver connected to a laptop or tablet}
  \label{fig:lucibel}
\end{figure}

\section{Performances and Outlook}

Lucibel has developed with its partner PureLiFi a fully functional commercial LiFi luminaire for enterprise applications, commercialized in its current version in September 2016. This is the first LiFi luminaire on the market.

The dimmable Lucibel Ores LiFi downlight \cite{oreslifi} implements proprietary technologies to perform energy and spectrum efficient modulation of a standard white light LED for the downlink, and an IR LED for the uplink. The bidirectional link is rated up to 42 Mbps (megabits per second) and up to 8 users can connect per AP, within a very a cell of diameter 3 m for a ceiling height of 2.5 m. Whereas a WiFi acces point covers a much larger area, many users have to share the bandwidth and end up with a lower speed. 

Dimming is possible while a user is communicating through an AP and communications are encrypted with standard WPA2 authentication protocols. 

Characteristics such as the receiver’s form factor and the handover protocol from AP to AP will be improved in the second generation to be released in 2018, with a new energy efficient electronic architecture.  Other configurations such as 2 ft x 2 ft panels and floor lamps will be put on the market to broaden the range of installations and use cases. An updated LiFi API will improve network configuration and management and make the LiFi system a part of Lucibel’s cloud based lighting management system.

The LiFi luminaire is compatible with Power over Ethernet (PoE) technology. This key characteristic permits to transmit through a single RJ45 cable at the same time data and power, thus minimizing the necessary wiring for the deployment of a LiFi network infrastructure and ultimately reducing the installation costs. PoE architectures are also more energy efficient.   

All components of the LiFi system have an impact on the performance in terms of data rate and energy consumption but, once the best available modulation technology is implemented and interferences and spatial effects are controlled in laboratory conditions, the most important contributor to performance is the modulation response of the LED. A lower modulation frequency translates into a lower data rate, although the relation is not straightforward and published data are not easily comparable. Equalization, for one, improves the LED 3dB cutoff frequency but at the detriment of energy efficiency. Various advanced processing methods can boost the throughput but they are not necessarily easily implemented in a compact and cost effective device.

\begin{table}[h]
\begin{center} {
\begin{tabular}{lllll}
\hline
 & \multicolumn{1}{l}{Technology} & \multicolumn{1}{l}{Bit Rate} & \multicolumn{1}{l}{Reference}\\
\hline
& Phosphor coated blue LED & 1.1 Gbps & \cite{khalid20121} \\
& Color converted blue \selectlanguage{greek}m\selectlanguage{english}LED & 1.7 Gbps & \cite{chun2014visible} \\
& Multicolor R(Y)GB LED and \selectlanguage{greek}m\selectlanguage{english}LED & 2 -- 11.3 Gbps & \cite{cossu20123} \cite{chun2016led} \cite{wang20158} \\
& Color Converted Laser Diodes & 1 - 4 Gbps & \cite{lee2017gigabit} \cite{chi2017violet} \\
\hline
\end{tabular} }
\end{center}
\caption{Data transmission speeds reported for various white light solid-state lighting technologies. With the exception of phosphor coated blue LEDs which constitute the building block of current commercial LED luminaires, the other technologies are in the demonstration phase for lighting applications.}
\label{turns}
\end{table}

Commercial blue LEDs coated with a yellow phosphor to produce white light have modulation bandwidths limited to a few MHz due to the long photoluminescence lifetimes of the phosphors. The application of a blue filter at the receiver removes the slow component from the signal and enables modulation frequencies of up to 20 MHz \cite{ayyash2016coexistence}.  This is the standard technology that is used in Lucibel’s Ores LiFi luminaire. Several strategies are being pursued to overcome this limitation.

Micro-LEDs can offer optical modulation bandwidths in excess of 600 MHz thanks to their small active areas enabling high current density injection \cite{islim2017towards}. They can be positioned in arrays to enable parallel communication, which is of interest for communication with displays.

Another way to create white light is to mix optically the emissions of three Red, Green and Blue LEDs or micro-LEDs. For LiFi, this has two main advantages: there is no bandwidth limiting color converter coating and the three LEDs can be modulated separately, thus allowing for the transmission of three parallel independent information streams \cite{cossu20123}.

But the best way to improve the modulation response is to use GaN Laser Diodes, because their modulation speed is controlled by the photon lifetime, on the order of ps, instead of the carrier lifetime for LEDs, several orders of magnitude higher. A 2.6 GHz modulation frequency has been reported \cite{lee20154} and 15 Gbps (gigabits per second) data rate over a single blue channel \cite{viola201715}. While Laser Diodes are as of today an expensive option for indoor lighting, their cost is expected to decrease exponentially over the next five years because of mass market demand for applications such as automotive lighting, projectors and LIDAR for autonomous vehicle.

All these technologies have been rated in the laboratory for white light emission, intensity modulation and data transfer to a photodiode. The feasibility of over Gbps LiFi data transmission is established, auguring well for transition to product development and commercialization shortly. With Laser Diodes, data rates in excess of 100 Gbps are hypothesized \cite{tsonev2015towards}. Constant new developments in the fields of design and manufacturing of LEDs \cite{li2017localization}\cite{lozano2016metallic}, Laser Diodes \cite{lee2016dynamic}, photon converters \cite{leitao2017gb} and optical sensors \cite{lestoquoy2014multimaterial}\cite{fink2017lifi}, as well as in the field of signal processing \cite{zhang2017localisation}, are monitored closely by Lucibel's teams to identify and update the best available technology for its LiFi devices.

\section{Use Cases: Towards Large Scale Deployments}

VLC has been a subject of intense research \& development for more than 15 years with steady improvements in performance, cost, reliability and components' compactness. While many applications have been imagined, such as vehicle to vehicle communications or underwater data transmission, the exponential development of Solid State Lighting has directed the short-term developments towards the best defined and most valuable use cases. 

\begin{figure}[!h]
  \includegraphics[width=\linewidth]{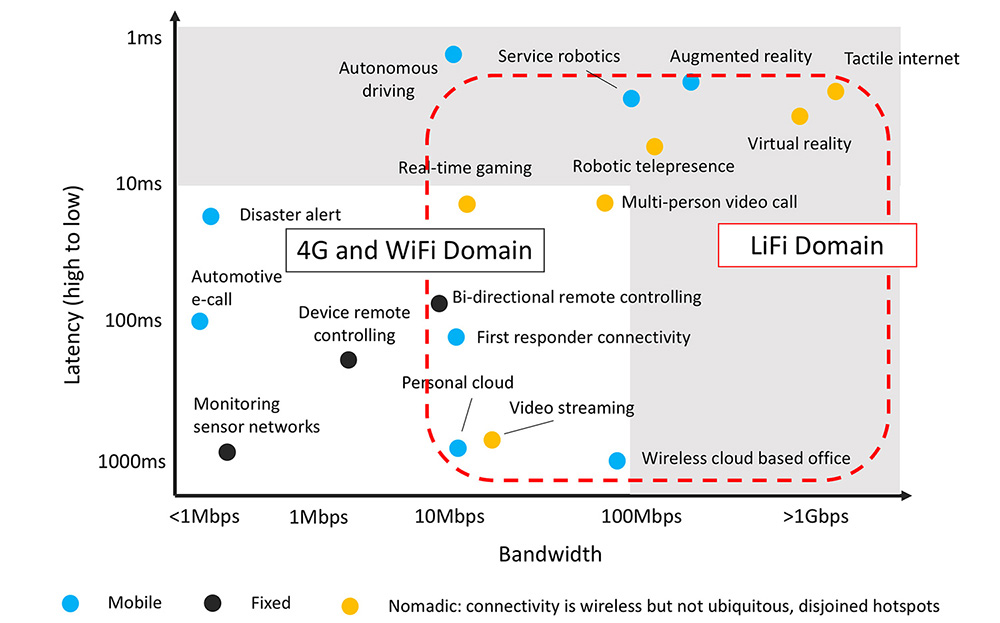}
  \caption{Projected data access demands of RF and LiFi technologies. Intrinsic low latency and the potential for a very large bandwidth make LiFi a technology of reference for video streaming and cloud based office as well as for emerging use cases such as Virtual Reality and applications of robotics in the industry, in the office or in public spaces. Sources: AT\&T \cite{att5G}, Lucibel}
  \label{fig:usecases}
\end{figure}

Lucibel has installed LiFi luminaires for more than 50 customers and has engaged extensively with them and communities, thus gaining a unique experience of the key value propositions for its first generation of products. Lucibel's customers are implementing LiFi as a powerful complement or alternative to WiFi and 4G, in environments where data exchange should be perfectly secure (banks, R\&D centers, defense, …), radio waves are not permitted or restricted (hospitals, pre-K schools, EMI sensitive industrial facilities such as natural gas compression stations) or connectivity should be guaranteed (conference rooms, hotels).  Train to train communications are also being explored. 

The selection of use cases is driven by the facts that, on the one hand, the light frequency range is interference free and not regulated and, on the other hand, Light Communications happen in the cone of light. In contrast to WiFi which suffers intrinsically from radiation leakage, the directionnality of light drastically limits the risk of eavesdropping and hacking of the network \cite{mostafa2015physical} \cite{chen2017physical}. Moreover, a key generation mechanism specific to OFDM can improve the internal communication security \cite{al2017secret}. 

Obvious limitations of the technology, such as the fact that the light has to be switched on with a minimum level of illumination, have to be acknowledged and constitute simple boundary conditions in the immense space of the indoor wireless use cases. 

There is no commercial deployment in the residential space yet. This relies on further cost reductions, miniaturization and integration of the receivers’ components in consumer electronics devices, expected to happen around 2020.

The data transmission speed of LiFi by Lucibel in the 10-50 Mbps range and the densification of Access Points position LiFi to be an enabler for the digital transformation of companies and infrastructures. Three numbers by Cisco explicit the global ever-accelerating need for bandwidth and wireless \cite{ciscovni2017}: by 2021 more than half of 17 billion connected devices will be mobile, 65\% of the IP traffic will be from mobile devices, 80\% of the internet traffic will be video requiring high speed wireless and the average RF wireless speed will be 20 Mbps. Mobile video streaming and personal cloud access are where LiFi excels and, with people in industrialized nations spending more than 90\% of their time indoors, lighting is poised to become a communications infrastructure of choice.

From an energy efficiency promotion perspective, with LiFi, lighting technologies offer enough value to customers for them to switch from incandescent and fluorescent lights to the much more energy efficient and cost competitive LEDs. This could be a considerable driver to increase LED adoption in existing buildings where first generation smart lighting’s reception has been so far tepid. The strong appetite for high quality and high speed internet access could push to make lighting infrastructure retrofits at the considerably quicker pace of IT infrastructure retrofits.

But what makes LiFi an even more exciting technology is the possibility to reach, with a next generation of LiFi by Lucibel devices, transmission and reception speeds in the 1 to 10 Gbps range with a very low latency inherent to optical technologies. New use cases will emerge.

Office robotic telepresence is a perfect example of synergy between the lighting and communication functions of Lucibel’s LiFi solution. In the workspace, lighting is ubiquitous and LiFi has a guaranteed coverage over the whole floor with a high minimum bandwidth, without interference and disconnection, perfectly adapted to the video streaming. Moreover LiFi, and more generally VLC technologies, have the inherent capacity to localize devices at the level of a luminaire, but even better, with a cm level precision thanks to signal processing \cite{do2016depth}\cite{lin2017experimental}. This is of tremendous value for many applications because indoor RF technologies are by far not as precise \cite{hu2016demonstration}. 

\begin{figure}[h]
  \includegraphics[width=\linewidth]{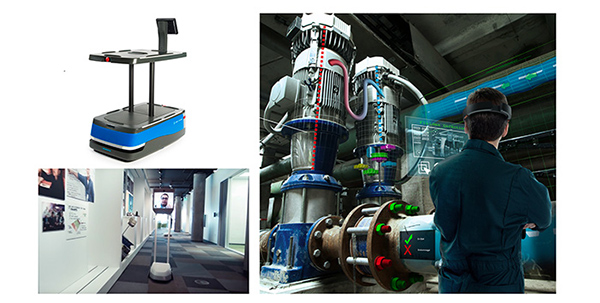}
  \caption{Emerging use cases for LiFi: warehouse robot, robotic telepresence, augmented reality. Pictures are for illustration purpose only. Sources: 6 River Systems, Suitable Technologies, Diota}
  \label{fig:newuse}
\end{figure}

LiFi's data transmission speed is beyond what's needed to connect IoT devices such as thermostats and presence detectors requiring at most 100 kbps. But the density of potential connections under a LiFi light spot is such that many devices can connect to a single AP and aggregate their bandwidth requirements. These devices can even harvest energy from the LED or Laser Diodes through the communication channel \cite{sandalidis2017illumination}\cite{passivevlc}, potentially solving the critical power supply issue for the IoT. Light communication directly between devices is also possible.

LiFi as a high-speed communication solution is extremely well positioned to be a solution of choice for the Industrial IoT, for example to feed and collect 3D data to and from AR devices at the service of workers on a manufacturing floor. Existing IoT architectures are highly centralized and heavily rely on a back-end core network for all decision-making processes. With LiFi, a large amount of data can be gathered in a secure way from multiple devices and processed at the edge by a local processor with the benefit of lower latency and reduced bandwidth requirements to communicate with the cloud \cite{bader2016front}. In the same spirit of decentralization, LiFi could be a crucial enabler for trusted transactions between IoT devices mediated by blockchains \cite{danzi2017analysis}.  

Lucibel is building with its partners, startups and other innovation driven companies, an ecosystem to bring progressively to the market the products that will fuel the large scale penetration of LiFi for these use cases and certainly many others. The first steps are to create awareness about the technology, to make its potential and also its limits known, to implement quickly the lessons learnt from early deployments and to train the workforce with the skills to build projects at the confluence of lighting and communication technologies. 

\hfill 
\break
\break
{\it Dr. C. Jurczak is Lucibel's Chief Scientific Officer, in Palo Alto, CA (USA).}

\medskip

\bibliographystyle{ieeetr}%Used BibTeX style is unsrt
\bibliography{LiFi}

\end{document}